\documentclass[prd,preprint,superscriptaddress,showpacs,byrevtex]{revtex4}
\newcommand{\be}{\begin{equation}}
\newcommand{\ee}{\end{equation}}
\newcommand{\bea}{\begin{eqnarray}}
\newcommand{\eea}{\end{eqnarray}}
\usepackage[dvips]{graphicx}
\begin{document}
\title{Screening and antiscreening in anisotropic QED \\ and QCD plasmas}
\author{Michael C. Birse} \email{mike.birse@man.ac.uk}
\affiliation{Theoretical Physics Group, Department of Physics and Astronomy, 
University of Manchester,
Manchester,M13 9PL, UK}
\author{Chung-Wen Kao} \email{kao@theory.ph.man.ac.uk}
\affiliation{Theoretical Physics Group, Department of Physics and Astronomy, 
University of Manchester,
Manchester,M13 9PL, UK}
\author{Gouranga C. Nayak} \email{nayak@shakti.lanl.gov}
\affiliation{T-8, Theoretical Division, Los Alamos National Laboratory,
Los Alamos, NM 87545, USA}

\date{\today}
\begin{abstract}
We use a transport-theory approach to construct the static propagator of 
a gauge boson in a plasma with a general axially- and reflection-symmetric 
momentum distribution. Non-zero magnetic screening is found if the 
distribution is anisotropic, confirming the results of a 
closed-time-path-integral approach. We find that the electric and magnetic 
screening effects depend on both the orientation of the momentum carried by 
the boson and the orientation of its polarization. In some orientations there 
can be antiscreening, reflecting the instabilities of such a medium. We 
present some fairly general conditions on the dependence of these effects 
on the anisotropy.

\end{abstract}  

\maketitle
%\narrowtext   
\newpage
\section{Introduction}

One of the important problems in non-equilibrium QED and QCD is
to study how such systems approach equilibrium.
In principle one should solve the Schwinger-Dyson equation to do
this at a purely quantum level, but it is impossible to do this without 
significant approximations \cite{sd}. For many practical purposes
one can use transport equations to study equilibration of QED and
QCD plasmas with particle collisions taken into 
account. For details we refer the reader to the extensive reviews
in the literature \cite{tr1, tr2}. However the scattering kernel in these 
transport equations can suffer from severe infrared divergences
due to ultra-soft photons or gluons exchanged in $t$-channel.
 
It is well known that the collective modes in the QED plasma screen the 
long-ranged Coulomb force \cite{kap, leBellac}.
The screened potential then has a Yukawa form in the medium:
\be
V(r)~\propto ~\frac{e^{-m_{D} r}}{r},
\ee
where $m_D$ is the Debye screening mass.
As a result the infrared behaviour of the exchanged gauge bosons is improved.
This Debye screening effect can be generalized to non-equilibrium QCD 
systems \cite{deb}
 and used in the transport equations describing quark-gluon 
plasmas \cite{deb1}.
 However for magnetic interactions the infrared problems persist.
In the case of QED in equilibrium the magnetic screening mass vanishes not 
just at one-loop level but to all orders in perturbation theory \cite{BIP}. 
It also has been argued that in QCD the magnetic screening mass should be
of the order of $g^2T$, and cannot be computed perturbatively \cite{leBellac}.

The situation is quite different in non-equilibrium systems, at least if 
they are anisotropic. Recently the closed-time-path-integral formalism was 
used to derive the magnetic screening mass at one-loop level \cite{fred}.
This showed that the magnetic screening mass in QED is non-zero if the
plasma has an anisotropic single-particle momentum distribution, 
{\it i.e.}~it depends on the direction of the momentum.
This would suggest that there is no natural infrared problem in 
such a system. 
More recently, however, Romatschke and Strickland \cite{RS} have used
the Hard Thermal Loop (HTL) approach to show how nonzero magnetic
screening effects can arise in anistopic plasmas. Their work shows that 
there can be antiscreening, in the sense of negative screening masses, as 
well as screening. This feature of out-of-equilibrium systems has also been 
noted by Arnold et al.~\cite{AMY}. It reflects the instabilities 
of these 
systems which have been studied by Mrowczynski \cite{mro,RM}. (For a review
of these ideas, see Ref.~\cite{ALM}.) The instability of these systems 
implies that, even though some components of the magnetic interactions are
screened, a simple perturbative treatment is still not adequate. 

It is well known that the polarization tensor of a gauge boson
derived from the HTL approach at leading order 
can also be obtained from a transport equation \cite{HTL,tr2}. 
The equivalence of the two approaches has also been generalized to the case 
of anisotropic but homogeneous systems \cite{mt}.
In this paper we use the transport-equation approach to derive the
same expression for the magnetic screening mass as was found in
\cite{fred}. The advantages of this approach are that the
physical picture is more transparent and its classical character is emphasized.
We present the general form of the propagator of a gauge boson in an anisotropic 
medium. This propagator is central to the scattering kernel of the 
transport equations \cite{deb1}. The structure of this propagator agrees
with that obtained by Romatschke and Strickland \cite{RS} who assumed a more
restricted functional form for the anisotropic distribution of particles
in the plasma. A minor technical difference from Ref.~\cite{RS} is that we 
work in Lorentz gauge rather than the temporal axial gauge, but we have
checked that our results are equivalent to theirs.

The propagator is more complicated than in the equilibrium case, 
since it depends on the the orientations of the field momentum and polarization 
relative to the axes of the anisotropy of the system. As a result the screening 
or antiscreening of a transverse magnetic field can itself be very anisotropic. 
Under some fairly general assumptions about the form of the anistropy, which are
likely to cover most cases of physical interest, including those studied in 
Ref.~\cite{RS}, we are able to derive conditions on the appearance of screening
or antiscreening in particular orientations.

The paper is organized as follows. In section II we present the derivation
of the polarization tensors in terms of the non-equilibrium
distribution function from transport equations.
In section III we discuss the screening of static fields and, in particular,
the form of the static polarization tensor for systems with cylindrical 
symmetry. We summarize our results and their implications in section IV.

\section{Derivation of the polarization tensor from transport equations}

The polarization tensor at one-loop level for a gauge boson in a plasma can 
be derived from classical transport equations \cite{HTL} even if the system 
is out of equilibrium \cite{mt}. As a simple example, consider first the case of a gas 
of electrons interacting with a weak, space-time dependent electromagnetic field. 
The single-particle distribution function $f({\bf x},{\bf p},t)$ obeys the 
Boltzmann equation,
\be
\frac{\partial f({\bf x},{\bf p},t)}{\partial t}+
{\bf v} \cdot \nabla_{x} f({\bf x},{\bf p},t)+
{\bf F}({\bf x},t) \cdot \nabla_{p} f({\bf x},{\bf p},t)=0,
\label{tr1}
\ee
where ${\bf v} ={\bf p}/{E({\bf p})}$ is the velocity of a particle with momentum 
{\bf p} and
\be
{\bf F}({\bf x},t) =e[{\bf E}({\bf x},t) + {\bf v} \times {\bf B}({\bf x},t)],
\ee
in terms of the electric field {\bf E} and the magnetic field {\bf B}.

We assume that on the distance- and time-scales of interest the plasma is 
close to homogeneous and so we can expand $f({\bf x},{\bf p})$ around an 
unperturbed distribution function $f_0({\bf p})$ as follows:
\be
f({\bf x},{\bf p},t)=f_0({\bf p})+f_1({\bf x},{\bf p},t).
\ee
This expansion is similar to the one which is usually used in the
derivation of the transport coefficient except that $f_0({\bf p})$ 
does not need to be the equilibrium Fermi-Dirac distribution.
The transport equation (\ref{tr1}) to first order in $f_{1}$ and
the electromagnetic fields is
\be
\frac{\partial f_1}{\partial t}+
{\bf v} \cdot \nabla_{x} f_1=-
{\bf F} \cdot \nabla_{p} f_0.
\ee
Solving this for $f_1$ yields 
\bea
f_1({\bf x},{\bf p},t)&=&-e \nabla_p f_0({\bf p}) \cdot \int^{t}_{-\infty} dt'
[{\bf E} (t',{\bf x}-{\bf v}(t-t'))
+{\bf v} \times {\bf B} (t',{\bf x}-{\bf v}(t-t'))] \nonumber \\
&=&-e \nabla_p f_0({\bf p}) \cdot \int^{\infty}_{0} d\tau
[{\bf E} (t-\tau,{\bf x}-\tau{\bf v})
+{\bf v} \times {\bf B} (t-\tau,{\bf x}-\tau{\bf v})].
\label{f1}
\eea

The induced current density is given by
\be
j_\mu^{ind}({\bf x},t)=e\int \frac{d^3p}{(2\pi)^3} v_{\mu} 
f_{1}({\bf x},{\bf p},t),
\label{in1}
\ee
where $v^\mu=p^{\mu}/E(\bf p)=(1,{\bf v})$.
Using the above expression for $f_1({\bf x},{\bf p},t)$, we find that the 
induced current can be expressed in terms of the fields as
\be
j_\mu^{ind} (x)=-e^2\int \frac{d^3p}{(2\pi)^3} \,v_{\mu}\,
\nabla_p f_0({\bf p}) \cdot \int^{\infty}_{0} d\tau 
[{\bf E} (t-\tau,{\bf x}-\tau{\bf v})
+{\bf v} \times {\bf B} (t-\tau,{\bf x}-\tau{\bf v})].
\ee
Taking the Fourier transform of this we get
\be
j_\mu^{ind} (k)=-ie^2\int \frac{d^3p}{(2\pi)^3}\, v_{\mu}\, 
\frac{\nabla_p f_0({\bf p})}{v \cdot k+i\epsilon} \cdot [{\bf E} (k)
+{\bf v} \times {\bf B} (k)]. 
\label{j1}
\ee
To get this result
we have regulated the $\tau$ integral by inserting a factor of $e^{-\epsilon\tau}$ 
with $\epsilon\rightarrow 0^+$ and we have used the fact that the Fourier 
transform of $\int_0^\infty d\tau e^{-\epsilon \tau} f(x-v\tau)$ is 
given by $if(k)/v  \cdot k +i\epsilon$. 
For isotropic systems, the gradient $\nabla_p f_0({\bf p})$ is proportional
to {\bf p}. The second term in Eq.~(\ref{j1}) vanishes in such cases and so there
is no response of the plasma to a magnetic field.

The Fourier components of the the fields can be expressed in terms of
the electromagnetic potential $A^\mu(k)$ as
\bea
{\bf B}(k)=i{\bf k}\times {\bf A}(k), \nonumber \\
{\bf E}(k)=-i{\bf k}A_{0}(k)+ik_{0} {\bf A}(k).
\eea
This allows us to express the current in the form
\bea
j_\mu^{ind} (k)&=&-ie^2\int \frac{d^3p}{(2\pi)^3} v_{\mu} 
\frac{\nabla_p f_0({\bf p})}{v \cdot k+i\epsilon} \cdot 
\left[-i{\bf k}A_{0}(k)+ik_{0} {\bf A}(k)+i({\bf v}\cdot{\bf A}(k)){\bf k}
-i({\bf v}\cdot{\bf k}){\bf A}(k)\right]
\nonumber \\
&=&-e^2\int \frac{d^3p}{(2\pi)^3}v_{\mu}\left[\frac{{\bf k}\cdot\nabla_{p}
f_{0}}{v\cdot k+i\epsilon}\,v\cdot A(k)
-\nabla_{p}f_{0}\cdot{\bf A}(k)\right]. \label{indcur}
\eea
We can introduce a polarization tensor, defined in terms of the induced current by
\be
j_\mu^{ind} (k)=\Pi_{\mu \nu} (k) A^\nu (k).
\label{j2}
\ee
The polarization tensor obtained from the induced current is of course only 
the matter part, arising from the gas of electrons. The full tensor also includes
a vacuum part which can be calculated using the standard field-theoretic methods.
After some algebra we get from Eq.~(\ref{indcur}) the matter polarization tensor
in the Lorentz gauge,
\bea
\Pi^{00}(k)&=&-e^{2}\int\frac{d^{3}p}{(2\pi)^{3}} 
\left[\frac{{\bf k}\cdot \nabla_{p}f_{0}({\bf p})}{k_{0}
-{\bf k}\cdot{\bf v}+i\epsilon}\right], \nonumber \\
\Pi^{0i}(k)&=&-e^2\int\frac{d^{3}p}{(2\pi)^{3}}
\left[\frac{{\bf k}\cdot \nabla_{p}f_{0}({\bf p})}{k_{0}
-{\bf k}\cdot{\bf v}+i\epsilon}\right]v_{i}, \nonumber \\
\Pi^{ij}(k)&=&e^2\int\frac{d^{3}p}{(2\pi)^{3}}
\left[\frac{f_{0}({\bf p})}{E({\bf p})}(\delta_{ij}-v_{i}v_{j})
-\frac{{\bf k}\cdot \nabla_{p}f_{0}({\bf p})}{k_{0}-{\bf k}\cdot{\bf v}+i\epsilon}
v_{i}v_{j}\right].
\label{nw}
\eea
If the particles are massless (or the typical energies are high enough that 
we may neglect their masses), the velocity {\bf v} may be replaced by the unit 
vector $\hat{\bf p}$ and the energy $E({\bf p})$ by $|{\bf p}|$.

To generalize this result to QCD, one should replace the space-time derivatives 
in Eq.~(\ref{tr1}) by covariant derivatives. However, so long as $gA_{\mu}^{a}$
is small, we can neglect the higher-order terms that this introduces. We can
then describe the QCD plasma by equations with the same form as Eq.~(\ref{tr1}).
The resulting contibutions to the gluon polarisation tensor have same forms
as for the photon except for a trivial factor which counts the relevant degrees 
of freedom. In QCD case one simply needs to replace $f({\bf p})$ in the expression
(\ref{nw}) for the polarization tensor by a sum of quark, antiquark and gluon 
terms \cite{BIP}, $2N_{f}(f_{q}({\bf p})+f_{\bar{q}}({\bf p}))+2N_{c}f_{g}({\bf p})$.
The factor of 2 before $N_{f}$ is due to the spin degree of freedom.

\section{Screening of static fields in an anisotropic plasma}

The long-distance behaviour of the static potential is governed by the low-momentum
behaviour of the static photon propagator. In the static limit, $k_0=0$, the 
polarization tensor Eq.~(\ref{nw}) depends the direction of {\bf k} only. 
Denoting the static tensor by $\bar{\Pi}^{\mu\nu}(\hat{\bf k})$, we find that its 
components take the forms
\bea
\bar{\Pi}^{00}(\hat{\bf k})&=&e^{2}\int\frac{d^{3}p}{(2\pi)^{3}}\, |{\bf p}|
\left[\frac{\hat{\bf k}\cdot\nabla_{p}f_{0}({\bf p})}{\hat{\bf k}\cdot{\bf p}
-i\epsilon}\right], \\
\bar{\Pi}^{0i}(\hat{\bf k})&=&e^2\int\frac{d^{3}p}{(2\pi)^{3}}\, |{\bf p}|
\left[\frac{\hat{\bf k}\cdot\nabla_{p}f_{0}({\bf p})}
{\hat{\bf k}\cdot{\bf p}-i\epsilon}\right]\hat{p}_{i},  \\
\bar{\Pi}^{ij}(\hat{\bf k})&=&e^2\int\frac{d^{3}p}{(2\pi)^{3}}
\left[\frac{f_{0}({\bf p})}{|{\bf p}|}(\delta_{ij}-\hat{p}_{i}\hat{p}_{j})
+|{\bf p}| \frac{\hat{\bf k}\cdot\nabla_{p}f_{0}({\bf p})}{\hat{\bf k}
\cdot{\bf p}-i\epsilon}
\hat{p}_{i}\hat{p}_{j}\right],\label{pibar}
\eea
in a plasma of massless particles.

In the familiar case of isotropic matter, the static polarisation tensor takes the 
form \cite{kap,leBellac}
\be
\bar\Pi^{\mu\nu}=\bar\Pi_L L^{\mu\nu}+\bar\Pi_T T^{\mu\nu},
\label{iso}
\ee
where, in the rest frame of the matter, $L^{\mu\nu}=-\delta_{\mu 0}\delta_{\nu 0}$
and $T^{\mu\nu}=-g^{\mu\nu}+k^\mu k^\nu/k^2-L^{\mu\nu}$. For comparison with the 
results for anisotropic matter, we note that the corresponding propagator
can be written
\be
D^{00}(0,{\bf k})={-1\over |{\bf k}|^2+m_D^2},\qquad
D^{ij}(0,{\bf k})={1\over|{\bf k}|^2+m_g^2}T^{ij}
+\alpha{\hat k_i\hat k_j\over |{\bf k}|^2},
\ee
where $\alpha$ is the gauge-fixing parameter ($\alpha=1$ in Feynman gauge) and
the Debye screening mass is, at the order to which we are working,
$m_D^2 = -{\rm Re}\left[\bar\Pi_{00}\right] 
= {\rm Re}\left[\bar\Pi_{L}\right]$.
Similarly, the magnetic screening mass is 
\be
m_g^2 = \frac{1}{2}T_{ij}{\rm Re}\left[\bar\Pi^{ij}\right]
= {\rm Re}\left[\bar\Pi_{T}\right],\label{magmass}
\ee
but this vanishes in isotropic matter as discussed above.

In anisotropic matter, the polarization tensor cannot be reduced to just two 
screening masses since it contains, in general, off-diagonal elements and it 
depends on the direction of the momentum carried by the field and on the 
direction of polarization vector. To examine this in more detail we make some 
specific assumptions about the form of the momentum distribution $f_0({\bf p})$, 
which we believe should be appropriate in the context of a relativistic 
heavy-ion collision. 
For matter near the collision axis, we assume that the system is symmetric about 
this axis. We therefore consider a cylindrically symmetric form for $f_0({\bf p})$. 
Also, if we work in the local rest-frame of the matter, there is no net flow and 
we can assume that the distribution is symmetric under reflections,
\be 
f_0(-{\bf p})=f_0({\bf p}). \label{eq:reflection}
\ee

In this case, as we shall see, the mixed space-time components of the tensor 
are purely
 imaginary. This means that
we can still define a Debye mass as above, although this mass now depends on 
direction. If we use Eq.~(\ref{magmass}) to define an averaged magnetic mass from 
the space components of the static tensor (\ref{pibar}), then we 
find that it
reproduces the result for the screening mass given in Ref.~\cite{fred} using the
more complicated closed-time-path-integral method,
\be
m_g^2(\hat {\bf k})=\frac{e^2}{2}\int \frac{d^3p}{(2\pi)^3}\left[
[1 +({\bf \hat p \cdot \hat k})^2]
\frac{f({\bf p})}{|{\bf p}|} 
+ [1 -({\bf \hat p \cdot \hat k})^2]
\,\frac{{\bf \hat k \cdot \nabla_{p}}f({\bf p})}{{\bf \hat k \cdot \hat p}}\right].
\label{mgef}
\ee
Note that the derivation of (\ref{mgef}) in \cite{fred} does not 
rely on the particular form of the distribution assumed here. 

Without loss of generality, we can choose the $z$-axis along our axis of 
symmetry and the momentum of the field to lie in the $xz$-plane. We can then 
express {\bf k} in the form
\be
{\bf k}=k_{\rho}\hat{\bf x}+k_{z}\hat{\bf z},
\ee
with $k_{\rho},\ k_{z}\geq 0$. In terms of integrals over the longitudinal
and radial components of {\bf p} and the angle $\phi$ between {\bf p}
and {\bf k} in the $xy$-plane, 
the components of the static polarization tensor can be written
\bea
\bar{\Pi}^{00}(\hat{\bf k})&=&\frac{e^2}{(2\pi)^{3}}\int^{\infty}_{-\infty}dp_z
\int^{\infty}_{0}p_\rho dp_\rho\int_{0}^{2\pi}d\phi
\sqrt{p_{\rho}^2+p_{z}^2}
\left[\frac{k_{\rho}\frac{\partial f_{0}}{\partial p_\rho}\cos\phi+k_{z}
\frac{\partial f_{0}}{\partial p_z}}{p_\rho k_{\rho}\cos\phi+p_{z}k_{z}-i\epsilon}
\right]
, \nonumber \\
\bar{\Pi}^{0x}(\hat{\bf k})&=&\frac{e^2}{(2\pi)^{3}}\int^{\infty}_{-\infty}dp_z
\int^{\infty}_{0}p_\rho dp_\rho\int_{0}^{2\pi}d\phi
\left[\frac{k_{\rho}\frac{\partial f_{0}}{\partial p_\rho}\cos\phi+k_{z}
\frac{\partial f_{0}}{\partial p_z}}{p_\rho k_{\rho}\cos\phi+p_{z}k_{z}-i
\epsilon}p_{\rho}\cos\phi
\right]
, \nonumber \\
\bar{\Pi}^{0z}(\hat{\bf k})&=&\frac{e^2}{(2\pi)^{3}}\int^{\infty}_{-\infty}dp_z
\int^{\infty}_{0}p_\rho dp_\rho\int_{0}^{2\pi}d\phi
\left[\frac{k_{\rho}\frac{\partial f_{0}}{\partial p_\rho}\cos\phi+k_{z}
\frac{\partial f_{0}}{\partial p_z}}{p_\rho k_{\rho}\cos\phi+p_{z}k_{z}-i
\epsilon}p_{z}\right]
, \nonumber \\
\bar{\Pi}^{xx}(\hat{\bf k})&=&\frac{e^2}{(2\pi)^{3}}\int^{\infty}_{-\infty}dp_z
\int^{\infty}_{0}p_\rho dp_\rho\int_{0}^{2\pi}d\phi
\frac{1}{\sqrt{p_{\rho}^{2}+p_{z}^2}}
\left[\frac{(p_z^2+p_\rho^{2}\sin^{2}\phi)f_{0}}{p_\rho^{2}+p_z^{2}}\right.
\nonumber \\
&&\qquad\qquad\qquad\qquad\qquad\qquad\qquad\qquad\qquad\;
\left.+\frac{k_{\rho}\frac{\partial f_{0}}{\partial p_\rho}\cos\phi+k_{z}
\frac{\partial f_{0}}{\partial p_z}}{p_\rho k_{\rho}\cos\phi+p_{z}k_{z}-i\epsilon}
p_{\rho}^{2}\cos^{2}\phi\right], \nonumber \\
\bar{\Pi}^{yy}(\hat{\bf k})&=&\frac{e^2}{(2\pi)^{3}}\int^{\infty}_{-\infty}dp_z
\int^{\infty}_{0}p_\rho dp_\rho\int_{0}^{2\pi}d\phi
\frac{1}{\sqrt{p_\rho^{2}+p_z^2}}
\left[\frac{(p_z^2+p_\rho^{2}\cos^{2}\phi)f_{0}}{p_\rho^{2}+p_z^{2}}\right.
\nonumber \\
&&\qquad\qquad\qquad\qquad\qquad\qquad\qquad\qquad\qquad\;
\left.+\frac{k_{\rho}\frac{\partial f_{0}}{\partial p_\rho}\cos\phi+k_{z}
\frac{\partial f_{0}}{\partial p_z}}{p_\rho k_{\rho}\cos\phi+p_zk_{z}-i\epsilon}
p_\rho^{2}\sin^{2}\phi\right], \nonumber \\
\bar{\Pi}^{zz}(\hat{\bf k})&=&\frac{e^2}{(2\pi)^{3}}\int^{\infty}_{-\infty} dp_z
\int^{\infty}_{0}p_\rho dp_\rho\int_{0}^{2\pi}d\phi
\frac{1}{\sqrt{p_\rho^{2}+p_z^2}}
\left[\frac{p_\rho^{2}f_{0}}{p_\rho^{2}+p_z^{2}}\right.
\nonumber \\
&&\qquad\qquad\qquad\qquad\qquad\qquad\qquad\qquad\qquad\;
\left.+\frac{k_{\rho}\frac{\partial f_{0}}{\partial p_\rho}\cos\phi
+k_{z}\frac{\partial f_{0}}{\partial p_z}}
{p_\rho k_{\rho}\cos\phi+p_zk_{z}-i\epsilon}
p_z^{2}\right], \nonumber \\
\bar{\Pi}^{xz}(\hat{\bf k})&=&\frac{e^2}{(2\pi)^{3}}\int^{\infty}_{-\infty}dp_z
\int^{\infty}_{0}p_\rho dp_\rho\int_{0}^{2\pi}d\phi
\frac{p_\rho p_z}{\sqrt{p_\rho^{2}+p_z^2}}\left[
\frac{k_{\rho}\frac{\partial f_{0}}{\partial p_\rho}\cos\phi+k_{z}
\frac{\partial f_{0}}{\partial p_z}}
{p_\rho k_{\rho}\cos\phi+p_zk_{z}-i\epsilon}\cos\phi\right], \label{eq:stpol}
\eea
The expressions for the $0y$, $xy$ and $yz$ components have been omitted because 
their integrands are odd functions of $\phi$ and so they integrate to zero.

In general, with $k_{\rho} \neq 0$ and $k_z \neq 0$, we can define 
$\gamma = k_z/k_\rho$ and carry out the integration over $\phi$ to get 
\bea
\bar\Pi^{00}&=&-\,\frac{e^2}{\pi^2}\int^{\infty}_{0}dp_z
\int^{\infty}_{0}dp_{\rho}\frac{p_{\rho}}{\sqrt{p_{z}^{2}+p_{\rho}^{2}}}\, f_{0} 
\nonumber \\
&&+\, \frac{e^2}{2\pi^2}\int^{\infty}_{0}dp_z\int_{0}^{\infty}dp_{\rho}
\frac{p_z}{\sqrt{p_{z}^{2}+p_{\rho}^{2}}}
\left[p_{\rho}\frac{\partial f_{0}}{\partial p_{z}}
-p_{z}\frac{\partial f_{0}}{\partial p_{\rho}}\right] \nonumber \\
&&+\,\frac{e^2}{2\pi^2}\int^{\infty}_{0}dp_z\int_{0}^{\gamma p_{z}}dp_{\rho}
\frac{\sqrt{p_{z}^{2}+p_{\rho}^{2}}}{p_{z}}
\left[p_{\rho}\frac{\partial f_{0}}{\partial p_{z}}
-p_{z}\frac{\partial f_{0}}{\partial p_{\rho}}\right]\frac{k_zp_z}
{\sqrt{k_{z}^{2}p_{z}^{2}-k_{\rho}^{2}p_{\rho}^{2}}}, \nonumber \\
%\bar\Pi^{00}&=&-\,\frac{e^2}{2\pi^2}\int^{\infty}_{0}dp_z
%\int^{\infty}_{0}dp_{\rho}\frac{p_{\rho}}{\sqrt{p_{z}^{2}+p_{\rho}^{2}}}\, f_{0}
%-\frac{e^2}{2\pi^{2}}\int^{\infty}_{0}dp_z p_zf_{0}(p_\rho=0,p_z) \nonumber \\
%&&+\,\frac{e^2}{2\pi^2}\int^{\infty}_{0}dp_z\int_{0}^{\gamma p_{z}}dp_{\rho}
%\frac{\sqrt{p_{z}^{2}+p_{\rho}^{2}}}{p_{z}}
%\left[p_{\rho}\frac{\partial f_{0}}{\partial p_{z}}
%-p_{z}\frac{\partial f_{0}}{\partial p_{\rho}}\right]\frac{k_zp_z}
%{\sqrt{k_{z}^{2}p_{z}^{2}-k_{\rho}^{2}p_{\rho}^{2}}}, \nonumber \\
\bar\Pi^{0z}&=&-
i\frac{e^2}{2\pi^2}\int^{\infty}_{0}dp_z\int^{\infty}_{\gamma p_{z}}dp_{\rho}
\left[p_{\rho}\frac{\partial f_{0}}{\partial p_{z}}
-p_{z}\frac{\partial f_{0}}{\partial p_{\rho}}\right]\frac{k_zp_z}
{\sqrt{k_{\rho}^{2}p_{\rho}^{2}-k_{z}^{2}p_{z}^{2}}}, \nonumber \\
\bar\Pi^{0x}&=&-\gamma\bar\Pi_{0z}, \nonumber \\
\bar\Pi^{xx}&=&-\,\frac{e^2}{2\pi^2}\int^{\infty}_{0}dp_z
\int^{\infty}_{0}dp_{\rho}\frac{\gamma^{2}p_{z}}{\sqrt{p_{z}^{2}
+p_{\rho}^{2}}}\left[p_{\rho}\frac{\partial f_{0}}{\partial p_{z}}
-p_{z}\frac{\partial f_{0}}{\partial p_{\rho}}\right] \nonumber \\
&&+\,\frac{e^2}{2\pi^2}\int^{\infty}_{0}dp_z\int_{0}^{\gamma p_{z}}dp_{\rho}
\frac{\gamma^{2}p_{z}}{\sqrt{p_{z}^{2}+p_{\rho}^{2}}}
\left[p_{\rho}\frac{\partial f_{0}}{\partial p_{z}}
-p_{z}\frac{\partial f_{0}}{\partial p_{\rho}}\right]\frac{k_zp_z}
{\sqrt{k_{z}^{2}p_{z}^{2}-k_{\rho}^{2}p_{\rho}^{2}}}, \nonumber \\
\bar\Pi^{yy}&=&\frac{e^2}{2\pi^2}\int^{\infty}_{0}dp_z
\int^{\infty}_{0}dp_{\rho}\frac{\gamma^{2}p_{z}}{\sqrt{p_{z}^{2}
+p_{\rho}^{2}}}\left[p_{\rho}\frac{\partial f_{0}}{\partial p_{z}}
-p_{z}\frac{\partial f_{0}}{\partial p_{\rho}}\right] \nonumber \\
&&-\,\frac{e^2}{2\pi^2}\int^{\infty}_{0}dp_z\int_{0}^{\gamma p_{z}}dp_{\rho}
\frac{\gamma^{2}p_{z}}{\sqrt{p_{z}^{2}+p_{\rho}^{2}}}
\left[p_{\rho}\frac{\partial f_{0}}{\partial p_{z}}
-p_{z}\frac{\partial f_{0}}{\partial p_{\rho}}\right]
\frac{\sqrt{k_{z}^{2}p_{z}^{2}-k_{\rho}^{2}p_{\rho}^{2}}}{k_{z}p_{z}}, \nonumber \\
%\bar\Pi^{zz}&=&\frac{e^2}{2\pi^2}\int^{\infty}_{0}dp_z
%\int^{\infty}_{0}dp_{\rho}\frac{p_{\rho}}{\sqrt{p_{z}^{2}+p_{\rho}^{2}}}\, f_{0}
%-\frac{e^2}{2\pi^{2}}\int^{\infty}_{0}dp_z p_zf_{0}(p_\rho=0,p_z) \nonumber \\
%&&+\,\frac{e^2}{2\pi^2}\int^{\infty}_{0}dp_z\int_{0}^{\gamma p_{z}}dp_{\rho}
%\frac{p_{z}}{\sqrt{p_{z}^{2}+p_{\rho}^{2}}}
%\left[p_{\rho}\frac{\partial f_{0}}{\partial p_{z}}
%-p_{z}\frac{\partial f_{0}}{\partial p_{\rho}}\right]\frac{k_zp_z}
%{\sqrt{k_{z}^{2}p_{z}^{2}-k_{\rho}^{2}p_{\rho}^{2}}}, \nonumber \\
\bar{\Pi}_{zz}&=&-\frac{e^2}{2\pi^2}\int^{\infty}_{0}dp_z
\int^{\infty}_{0}dp_{\rho}\frac{p_{z}}{\sqrt{p_{z}^{2}
+p_{\rho}^{2}}}\left[p_{\rho}\frac{\partial f_{0}}{\partial p_{z}}
-p_{z}\frac{\partial f_{0}}{\partial p_{\rho}}\right] \nonumber \\
&&+\,\frac{e^2}{2\pi^2}\int^{\infty}_{0}dp_z\int_{0}^{\gamma p_{z}}dp_{\rho}
\frac{p_{z}}{\sqrt{p_{z}^{2}+p_{\rho}^{2}}}
\left[p_{\rho}\frac{\partial f_{0}}{\partial p_{z}}
-p_{z}\frac{\partial f_{0}}{\partial p_{\rho}}\right]
\frac{k_{z}p_{z}}{\sqrt{k_{z}^2p_{z}^2-k_{\rho}^{2}p_{\rho}^{2}}} \nonumber \\
\bar\Pi^{xz}&=&\frac{e^2}{2\pi^2}\int^{\infty}_{0}dp_z
\int^{\infty}_{0}dp_{\rho}\frac{\gamma p_{z}}{\sqrt{p_{z}^{2}+p_{\rho}^{2}}}
\left[p_{\rho}\frac{\partial f_{0}}{\partial p_{z}}
-p_{z}\frac{\partial f_{0}}{\partial p_{\rho}}\right] \nonumber \\
&&-\,\frac{e^2}{2\pi^2}\int^{\infty}_{0}dp_z\int_{0}^{\gamma p_{z}}dp_{\rho}
\frac{\gamma p_z}{\sqrt{p_{z}^{2}+p_{\rho}^{2}}}
\left[p_{\rho}\frac{\partial f_{0}}{\partial p_{z}}
-p_{z}\frac{\partial f_{0}}{\partial p_{\rho}}\right]\frac{k_zp_z}
{\sqrt{k_{z}^{2}p_{z}^{2}-k_{\rho}^{2}p_{\rho}^{2}}}.
\eea
Note that if the distribution function satisfies the reflection condition 
(\ref{eq:reflection}) then the time-time and space-space components in the 
static polarization tensor are real and the mixed space-time components
elements are purely imaginary. (This can be seen by making the change of variables
$p_{z}\rightarrow -p_{z}$, $\phi \rightarrow \phi+\pi$ in (\ref{eq:stpol})). 

The elements of this tensor do not have definite signs and so there can be
antiscreening in some directions as well as screening in others. Also, the 
appearance of a nonzero space-time component means that in general the Debye
screening of electric fields cannot be separated from the effects on magentic
fields. The structure of this tensor agrees with that obtained by Romatschke
and Strickland \cite{RS} in the temporal axial gauge. However the Lorentz gauge
used here is more convenient for handling static electric fields.

For our chosen form of distribution, the spatial part of static polarization tensor 
is real and symmetric. 
Hence the tensor $\bar\Pi^{ij}$ has three real, orthogonal eigenvectors. The 
condition that the polarization tensor be transverse reduces to 
\be
k_i\bar\Pi^{ij}=0,
\ee
in the static limit. This shows that {\bf k} is an eigenvector with zero eigenvalue.
By inspection of Eq.~(\ref{eq:stpol}) we see that the $y$-direction, perpendicular
to the plane containing {\bf k} and the collision axis, is also an eigenvector.
Hence three unit eigenvectors for the static tensor are
\be
\hat{\bf e}_{1}=\hat{\bf k},\quad \hat{\bf e}_{2}=\hat{\bf y},\quad \hat{\bf e}_{3}
=\hat{\bf k}\times\hat{\bf y}.
\ee 
We denote the corresponding eigenvalues of the spatial part of the tensor 
by $\bar{\Pi}_{1}=0$, $\bar{\Pi}_{2}$ and $\bar{\Pi}_{3}$.
In the coordinate system defined by these axes the spatial part of the tensor is 
diagonal. Also we find that $\bar{\Pi}^{01}=\bar{\Pi}^{02}=0.$ The 
corresponding static propagator for the gauge boson has the non-zero elements
\bea
&&\left(\begin{array}{cc}
D^{00}(0,{\bf k}),& D^{03}(0,{\bf k})\\
D^{30}(0,{\bf k}),& D^{33}(0,{\bf k})
\end{array}\right)
=\frac{-1}{(|{\bf k}|^{2}+\bar{\Pi}_{+})(|{\bf k}|^2+\bar{\Pi}_{-})}
\left(\begin{array}{cc}
|{\bf k}|^2+\bar{\Pi}_{3},& \bar{\Pi}^{03}\\
\bar{\Pi}^{03},& -|{\bf k}|^{2}+\bar{\Pi}^{00}
\end{array}\right), \nonumber \\
&&D^{11}(0,{\bf k})={1+\alpha\over|{\bf k}|^2},\qquad\qquad\qquad
D^{22}(0,{\bf k})={1\over|{\bf k}|^2+\bar{\Pi}_{2}({\bf k})}.
\eea
Here we have introduced the two eigenvalues $\bar\Pi_{\pm}$ of the full tensor 
in the time-3 subspace. These are given by 
\be
\bar\Pi_\pm= -\frac{1}{2}(\bar{\Pi}_{00}-\bar{\Pi}_{3})
\pm\frac{1}{2}\sqrt{(\bar{\Pi}_{00}+\bar{\Pi}_{3})^2
+4|\bar{\Pi}_{03}|^2} .
\ee

These results show that the screening or antiscreening of static electromagnetic 
fields in an anisotropic plasma can depend on both the orientation of {\bf k}, the 
momentum carried by the field, and the orientation of the field itself. 
It is worth considering two special cases. First, when the momentum carried
by the field is perpendicular to the collision axis ($k_z=0$ and hence 
$\hat{\bf e}_{3}=\hat{\bf z}$) we have
\be
\bar{\Pi}_{1}=\bar{\Pi}_{2}=0,\,\,\,\,\bar{\Pi}_{3}
=\bar{\Pi}^{zz},\,\,\,\bar{\Pi}_{03}=0.
\ee
In this case fields in the $y$-direction are unscreened. In contrast when the 
momentum carried by the field lies along the axis ($k_\rho=0$ and hence 
$\hat{\bf e}_{3}=\hat{\bf x}$) we have
\be
\bar{\Pi}_{1}=0,\,\,\,\bar{\Pi}_{2}=\bar{\Pi}_{3}=\bar{\Pi}^{xx},\,\,\,
\bar{\Pi}_{03}=0.
\ee
In this case the screening masses for the transverse components of the 
field are degenerate.
For both of these special directions of the momentum, there is no mixed 
time-3 component and so $-\bar\Pi_{00}$ and $\bar\Pi_3$ are both eigenvalues.

In general there can be antiscreening as well as screening of electromagnetic fields
in an anisotropic plasma. As discussed in Ref.~\cite{RS}, this reflects the fact that
such as system is not in equilibrium and hence is unstable. The instability of
a particular mode depends on the detailed form of the momentum distibution 
$f_0({\bf p})$. It is not possible to draw general conclusions about the pattern of 
screening masses except in cases where the combination of derivatives
\be
p_{\rho}\frac{\partial f_{0}}{\partial p_{z}}
-p_{z}\frac{\partial f_{0}}{\partial p_{\rho}}
\ee
has the same sign for all momenta. This covers many smooth anisotropic distributions,
including the examples studied in
 Ref.~\cite{RS}. 
A negative value for this 
combination corresponds to an oblate
 momentum distribution with 
$\langle p_z^2\rangle<\frac{1}{2}\langle p_\rho^2\rangle$,
and a positive value to a prolate distribution.
If the combination of derivatives has a definite sign, the integrals in 
$\bar\Pi_2=\bar\Pi_{yy}$ can be rewritten in a form which shows that they too have 
the same sign. For an oblate distribution we get $\bar\Pi_2<0$, indicating 
antiscreening while for a prolate one we get $\bar\Pi_2>0$ and screening. 

In the oblate case we also find that $\bar\Pi_{00}-\bar\Pi_3$ can be
written in a form which shows that it is negative definite. This implies that 
$\bar\Pi_+>0$ and hence there is always screening for one component of the field.
In addition, at $k_z=0$ we find that $\bar\Pi_{00}+\bar\Pi_3$ is negative and
$\bar\Pi_3$ is positive, and hence $\bar\Pi_-=\bar\Pi_3$ is positive. Since 
at $k_\rho=0$ we have $\bar\Pi_-=\bar\Pi_3=\bar\Pi_2<0$, this implies that 
$\bar\Pi_-$ must change from antiscreening to screening as the orientation 
of {\bf k} moves away from the collision axis. 

For prolate distributions, in contrast, the sign of $\bar\Pi_{00}-\bar\Pi_3$ is 
not determined and so it is not possible to make such definite statements
about $\bar\Pi_\pm$. Nonetheless the nonderivative term in $\bar\Pi_{00}$ (which is 
responsible
 for the usual Debye screening) is always negative and so $\bar\Pi_{00}$
and $\bar\Pi_{00}-\bar\Pi_3$ remain negative unless the distribution is  
strongly prolate. Hence we expect the eigenvalue $\bar\Pi_+$ to correspond to 
screening for all except the the most extreme prolate anisotropies.  
At $k_z=0$ we again find that $\bar\Pi_-=\bar\Pi_3$, but now this 
must be negative. At $k_\rho=0$ we have $\bar\Pi_3=\bar\Pi_2>0$ and so 
both $\bar\Pi_+$ and $\bar\Pi_-$ are positive for orientations of 
{\bf k} close to the axes of distributions that are not too strongly polate.
However, as the deformation increases, $\bar\Pi_{00}$ becomes less negative 
and so the range of orientations for which $\bar\Pi_-$ is positive shrinks. 
For strong enough deformations, $\bar\Pi_-$ can become negative for all 
orientations.

The existence of at least one unstable component of the field, for some
orientations of {\bf k}, agrees with the general results of Ref.~\cite{ALM}.
All of the general features
 of $\bar\Pi_2$ and $\bar\Pi_\pm$ discussed above can 
be seen in the numerical examples studied in Section V
 of Ref.~\cite{RS}, and 
also in the analyses
 of ideally planar or linear distributions in Ref.~\cite{ALM}.

\section{Conclusion}

In this paper we have used a transport-theory approach to derive the polarization 
tensor of a gauge boson in a plasma which is out of thermal equilibrium. 
We find that the magnetic screening mass at lowest order is non-zero as long as the 
single-particle distribution function is anisotropic, in contrast
to the more familiar case of a plasma in equilibrium. This confirms results 
previously found using a closed-time-path-integral approach.

The full propagator for static magnetic fields in such a medium has a complicated 
tensor structure and its eigenvalues need not be positive. There can thus be
antiscreening rather than screening for some components of the field.
We have considered in detail the case of a plasma with a cylindrically 
symmetric momentum distribution. Such a distribution is expected to be relevant
to relativistic heavy-ion collisions before thermal equilibrium has been reached,
where the axis of collision can provide a symmetry axis for the anisotropy.
In this case the spatial part of the  static polarization tensor is real 
and has three orthogonal principal axes:
$\hat{\bf k}$, lying along the direction the momentum carried by the gauge boson,
$\hat{\bf y}$, orthogonal to the plane of {\bf k} and the collision axis, and
$\hat{\bf k}\times\hat{\bf y}$. The last two of these have nonzero screening in 
general. In additional there can be an imaginary off-diagonal component mixing
the $\hat{{\bf k}}\times\hat{{\bf y}}$ and time directions.
In the special case where the momentum {\bf k} is perpendicular to the 
collision axis, there is no screening for the component in the $y$-direction.

We have also been able to derive various conditions on the signs of two of the 
eigenvalues of the tensor, under the assumption that $p_{\rho}\frac{\partial f_{0}}
{\partial p_{z}}-p_{z}\frac{\partial f_{0}}{\partial p_{\rho}}$ has a definite sign.
In particular we find that fields in the $y$-direction are screened if the
anisotropy is prolate, but antiscreened if it is oblate. We also find that there is 
one component which is always screened unless the distribution is extremely
prolate. 

These differences in the screening of interactions in different directions could
have important effects in the equilibration of the matter produced in relativistic 
heavy-ion collisions. Although only the screening or antiscreening of static fields 
has been examined in detail here, it will also be very interesting to explore dynamical 
aspects, such as damping rates and unstable modes, in these anisotropic systems 
\cite{RS,mro,RM,ALM}. 

\acknowledgements{We thank Fred Cooper and Emil Mottola
for useful discussions. MCB and CWK 
acknowledge support from the UK EPSRC. 
This research is also supported by the Department
of Energy, under contract W-7405-ENG-36.}

\end{document}